# Quantifying impacts of short-term plasticity on neuronal information transfer

Pat Scott,[1] Anna I. Cowan[2] and Christian Stricker[2,3]

[1]Department of Physics, McGill University, 3600 rue University, Montréal, QC H3A 2T8, Canada. patscott@physics.mcgill.ca

[2]Department of Neuroscience, The John Curtin School of Medical Research, The Australian National University, Canberra ACT 0200, Australia.

[3]ANU Medical School, The Australian National University, Canberra ACT 0200, Australia. christian.stricker@anu.edu.au

## Abstract

Short-term changes in efficacy have been postulated to enhance the ability of synapses to transmit information between neurons, and within neuronal networks. Even at the level of connections between single neurons, direct confirmation of this simple conjecture has proven elusive. By combining paired-cell recordings, realistic synaptic modelling and information theory, we provide evidence that short-term plasticity can not only improve, but also reduce information transfer between neurons. We focus on a concrete example in rat neocortex, but our results may generalise to other systems. When information is contained in the timings of individual spikes, we find that facilitation, depression and recovery affect information transmission in proportion to their impacts upon the probability of neurotransmitter release. When information is instead conveyed by mean spike rate only, the influences of short-term plasticity critically depend on the range of spike frequencies that the target network can distinguish (its effective dynamic range). Our results suggest that to efficiently transmit information, the brain must match synaptic type, coding strategy and network connectivity during development and behaviour.

# 1   Introduction

The primary means by which information moves about a brain or neural network is the recursive generation of action potentials (APs) in networks of synaptically connected neurons. How effectively information passes from one cell to another is seen in how much information the APs generated in a postsynaptic cell contain about the APs in the presynaptic cell. This is governed by the processes that lead APs in one cell to evoke APs in another. Here, we focus on the impact of the properties of the presynaptic terminal upon neuronal information transfer in neocortex.

As basic network components, neurons consist of an excitable cell membrane along which an electrical signal can be carried, and synapses by which such an excitation in one cell can invoke or suppress similar excitations in nearby cells. An electric potential difference is maintained across the cell membrane by channel proteins that transport ions between the intracellular and extracellular solutions. This potential is altered by the opening or closing of ion channels in the cell membrane. Such potential changes may be caused by the binding of a neurotransmitter, which typically is emitted from nerve terminals, to the cell membrane. Changes in the membrane potential also trigger further alterations in the activation state of channels maintaining the potential itself, generating a localised inversion of the potential (an action potential). This AP propagates along the neuronal membrane until it reaches a nerve terminal, where it may evoke the release of one or more synaptic vesicles containing neurotransmitter. The neurotransmitter diffuses through the extracellular space, and may bind to the membranes of other neurons.

Synapses show a broad range of activity-dependent adaptive behaviours [1-4]. Those that occur on the shortest time scales are known as short-term plasticity. Four types of short-term plasticity are commonly seen in neocortical synapses. Vesicle-depletion depression (VDD) reduces the probability that the arrival of an action potential at a synapse will cause neurotransmitter release, and occurs because successful releases deplete the pool of readily releasable synaptic vesicles [4-7]. Release-independent depression (RID) produces a similar reduction in the probability of vesicle release, but occurs regardless of whether previous APs have successfully evoked neurotransmitter release or not [1,3,8-10]. Although it has been argued that RID is a selection effect caused by stochastic state-changes of the release machinery [11], previous work suggests that RID indeed has a distinct physiological basis [3]. For this paper, only the phenomenological occurrence of RID is important, not its exact cause.

Facilitation (FAC) increases the probability of release, as vesicles or the release machinery are primed by a release-independent mechanism following AP arrival [2,4,6,10]. Frequency-dependent recovery (FDR) also increases release probability by reducing the time a synapse takes to recover from depression [3,12,13]. FDR is thought to be associated with recovery from RID rather than VDD [3].

Short-term plasticity is expected to affect information transmission and encoding in neural networks [3,5,6,14-16], with some suggesting that it might actively enhance the ability of synapses to transmit information [3,15,16]. Surprisingly, this key postulate has yet to be tested in any reliable, quantitative way. We fill this gap by obtaining recordings of connected pairs of neurons in rat neocortex, determining the biophysical properties of the synapses involved, and directly estimating information transfer between them. Specifically, we elucidate the individual impacts of VDD, RID, FAC and FDR upon information transfer, and how these effects might change under different coding and network schemes.

We find that short-term plasticity can both increase and decrease information transfer between cells. We see deviations from a simple proportionality between information transfer and release probability with changes in either the neural code or the neural network. Surprisingly, both effects can be traced to a reduction in the effective range of spike rates that can be distinguished by postsynaptic cells.

## 2     Methods

Experimentally manipulating a single aspect of short-term plasticity in a living system, and then obtaining the long periods of recordings required for an accurate estimate of its effect upon information transfer, is practically impossible. We therefore perform our information-transfer experiments using computational models, but extract parameter values for these models from experimental recordings of neuronal pairs exhibiting short-term plasticity.

Because we focus exclusively on the impact of short-term plasticity, we include detailed presynaptic dynamics in our models but neglect postsynaptic details, such as dendritic integration and modulation of intrinsic excitability. This approach naturally complements previous work on the impact of postsynaptic properties upon synaptic information transfer [17], where the influence of presynaptic plasticity was essentially ignored in favour of detailed postsynaptic models.

Because the relative importances of spike rates and spike timings to neural coding is still a matter of debate, we differentiate the impacts of short-term plasticity on these two coding schemes [18-25]. We focus on information carriage by excitatory synapses here, but inhibitory connections and analogue coding [26-28] would make for interesting future work; see e.g. [29] for an investigation of the information-passing abilities of networks utilising both analogue (electrical) and AP-based (chemical) signalling.

### 2.1  *Paired cell recordings.*

We recorded from connected pairs of excitatory neurons in layers IV and V of rat somatosensory cortex. We evoked stimulus sequences of APs in presynaptic cells under current clamp using short DC current pulses (5 ms, 1 nA), and recorded excitatory postsynaptic currents (EPSCs) from postsynaptic cells under voltage clamp.[1] We used constant rate and Poisson stimulus sequences, with characteristic frequencies of 5, 10, 20, 30, 35, 40 or 50 Hz. Constant-rate stimuli consisted of either 6 or 20 pulses, and all sequences were followed by a recovery pulse after a delay of 400 or 500 ms. Including recovery pulses, stimulus trains were 0.8 – 2.5 s long, with 15 s between stimuli. Not all stimulus combinations could be presented to each pair, due to the finite duration of experimental recordings. Each stimulus we presented to a cell was repeated 45–50 times. In some cases we randomised the order in which different stimuli were presented, in other cases we completed all 45-50 repeats before changing to a different stimulus. We averaged EPSCs for each stimulus and calculated their variances (**Fig. 1**). Further details of the recordings and the tissue preparation can be found in [9]. We identified a set of connections ($N = 11$) that covered the full range of observed short-term dynamics and possessed the best data quality.

---

1  Under the current clamp protocol, a cell's membrane potential is allowed to vary, and the stimulus current is provided via the recording electrode. Under voltage clamp, the membrane potential is held fixed and the current required to keep the cell at this potential is monitored via the recording electrode.

*2.2 Model Synapse.*

Based on the ideas of Tsodyks and Markram [3,5,7,15] and with inspiration from Dittman [2,12], we developed a holistic model of a synapse able to exhibit VDD, RID, FAC and FDR. Each of these processes is essentially described by two parameters: strength (S) and persistence in time ($\tau$). The model succinctly encapsulates the effects of all four processes, allowing them to occur simultaneously. Apart from the observed correlation between RID and FDR [3], the four processes occur independently of each another. This reflects the fact that whilst the underlying biophysical mechanisms giving rise to RID and FDR are thought to be shared, VDD and FAC are thought to be mechanistically distinct from the other forms of short-term plasticity.

When averaged over stochastic release events, the model synapse describes the average peak EPSC amplitudes elicited in response to a series of presynaptic spikes, as well as the variances of those EPSCs. We refer to this as the 'deterministic form' of the model. The model is defined by a set of four ordinary differential equations

$$\frac{dP_V}{dt} = \frac{1 - P_V}{\tau_{VDD}} - U_{SE} P_V \delta(t - t_{AP}), \tag{1}$$

$$\frac{dU_{FAC}}{dt} = \frac{U_0 - U_{FAC}}{\tau_{FAC}} + S_{FAC}(1 - U_{FAC})\delta(t - t_{AP}), \tag{2}$$

$$\frac{dU_{RID}}{dt} = \frac{1 - U_{RID}}{\tau_{RID}} - S_{RID} U_{RID} \delta(t - t_{AP}), \tag{3}$$

$$\frac{d\tau_{RID}}{dt} = \frac{\tau_0 - \tau_{RID}}{\tau_{FDR}} - S_{FDR} \tau_{RID} \delta(t - t_{AP}), \tag{4}$$

and three supplementary relations

$$U_{SE} = U_{FAC} U_{RID}, \tag{5}$$

$$I_{VClamp} = \sum_{\text{all connections}} A_{SE} U_{SE} P_V, \tag{6}$$

$$\sigma^2 = \frac{1}{n} A_{SE}^2 U_{SE} P_V (1 - U_{SE} P_V). \tag{7}$$

The values $t_{AP}$ are the arrival times of APs at the presynaptic terminal, and $\delta$ denotes the Dirac delta function. $P_V$ is the average probability of vesicle availability, and $U_{SE}$ is the probability that a vesicle would be released if an AP arrived at the synapse and found that a vesicle were available there for release (the 'utilisation of synaptic efficacy'). $U_{SE}$ is the product of $U_{FAC}$ and $U_{RID}$, the normalised amounts of release machinery actively facilitated at any time $t$ by FAC, and free from inactivation by RID, respectively (the "effective synaptic efficacy due to FAC/RID"). VDD, RID, FAC and FDR recover exponentially with persistences $\tau_{VDD}$, $\tau_{RID}$, $\tau_{FAC}$ and $\tau_{FDR}$. RID, FAC and FDR occur with respective normalised strengths $S_{RID}$, $S_{FAC}$ and $S_{FDR}$. The initial probability of vesicle release is $U_0$, and $\tau_0$ is the starting value of $\tau_{RID}$. $A_{SE} = nq$ is the 'absolute synaptic efficacy', the maximum postsynaptic current a single AP can elicit from the connection (when all vesicles are successfully released). Here n is the total number of vesicle release sites connecting the two cells, and q is the resulting quantal current evoked by a single vesicle. $A_{SE}$ is used to scale the normalised model to correctly reproduce the observed size of EPSCs. $I_{VClamp}$ is thus the total peak amplitude of the postsynaptic current elicited by multiple connections in response to presynaptic APs, and $\sigma^2$ is its variance.

The arrival of an AP causes an amount of neurotransmitter to be released proportional to both $P_V$ and $U_{SE}$ (**Eq. 6**). Because the postsynaptic cell is kept in voltage clamp, the resulting EPSC is directly proportional to the amount of neurotransmitter released; successful release at all sites would produce an EPSC of size $A_{SE}$. To implement VDD, $P_V$ is decremented in proportion to the amount of neurotransmitter released (**Eq. 1**), and allowed to relax exponentially back to its initial value (1.0) with persistence $\tau_{VDD}$. $U_{SE}$ is either incremented or decremented following AP arrival (**Eqs. 2, 3 and 5**), according to the relative strengths of RID and FAC ($S_{RID}$ and $S_{FAC}$). It then decays back to its initial value ($U_0$) with a competing exponential behaviour that depends upon the different recovery persistences of RID and FAC ($\tau_{RID}$, $\tau_{FAC}$). FDR is then implemented by decrementing $\tau_{RID}$ itself with each AP by the relative amount $S_{FDR}$ (**Eq. 4)**, which relaxes back exponentially to its initial value ($\tau_0$) with persistence ($\tau_{FDR}$).

The functional behaviour of the model is described by 9 parameters ($A_{SE}$, $U_0$, $S_{RID}$, $S_{FAC}$, $S_{FDR}$, $\tau_0$, $\tau_{VDD}$, $\tau_{FAC}$ and $\tau_{FDR}$), with one extra parameter (n) required to describe the variance. Between APs (i.e. when $t \neq t_{AP}$), the iterative solution to **Eqs. 1–4** is

$$P_{V,i+1} = 1 + (P_{V,i} - 1)e^{-dt/\tau_{VDD}}, \tag{8}$$

$$U_{FAC,i+1} = U_0 + (U_{FAC,i} - U_0)e^{-dt/\tau_{FAC}}, \tag{9}$$

$$U_{RID,i+1} = 1 + (U_{RID,i} - 1)\left(\frac{\tau_{RID,i}}{\tau_{RID,i+1}}\right)^{\tau_{FDR}/\tau_0} e^{-dt/\tau_0}, \tag{10}$$

$$\tau_{RID,i+1} = \tau_0 + (\tau_{RID,i} - \tau_0)e^{-dt/\tau_{FDR}}, \tag{11}$$

where d$t$ is the time step between point $i$ and point $i$+1. The existence of an exact solution drastically reduces the time required to produce simulated responses, allowing their information content to be measured within a tractable timeframe.

An example of the deterministic model's time evolution is given in **Fig. 2**. This figure gives a graphical illustration of the mathematical model's description of the dynamic impacts of VDD, RID, FDR and FAC upon release probability. For this example, the peak EPSC amplitudes have not been convolved with any time course of the current, and are represented simply by delta functions; for information transfer measurements, we instead used an exponential conductance time course (explained in Sec 3.4).

### 2.3   *Fitting of peak amplitudes of EPSCs and their variances.*

We used this deterministic model synapse to analyse recordings from our set of 11 connections (**Fig. 3**). Peak EPSC amplitudes were measured from local baselines, whilst a constant baseline due to recording noise was subtracted from the variance time courses. We simultaneously fitted these peak EPSC amplitudes and their variances obtained at different stimulus frequencies in $A_{SE}$, $U_0$, $S_{RID}$, $S_{FAC}$, $S_{FDR}$, $\tau_{VDD}$, $\tau_0$, $\tau_{FAC}$, $\tau_{FDR}$ and n using a minimised least-squares analysis. For fitting, EPSCs were typically given twice the weighting of variances, although poorer data quality meant that a weighting of 4:1 was used in one case, and variances were not included at all in two other cases. Fits appear poorer by eye for the first few pulses in each sequence of EPSCs only because those observations have larger experimental variances, so are weighted less strongly in the overall fit; the model in fact fits all stimuli equally well. Because we performed our information transfer calculations on a per-site basis (i.e. n = 1, see below), the only role for n was as an overall scaling factor for fitting the variances, allowing both means and variances of EPSCs to be used to simultaneously constrain the other 9 parameters.

It should be noted that the FDR parameters $S_{FDR}$ and $\tau_{FDR}$ only have rather subtle impacts upon the model output, so their values are largely unconstrained by the fitting procedure. Indeed, in some cases a comparable fit could be found with or without FDR. However, this reflects the numerical difficulty of exploring such a high-dimensional and complicated parameter space as much as it does the presence or absence of FDR. Many local maxima exist in the likelihood surface traversed in the least-squares analysis, and we are not entirely confident of having located the global best-fit in every case; to do so is a very difficult problem, and would require a dedicated study of its own. In any case, we know that FDR certainly exists in rat neocortex [3], so it must be included in our model and investigated in our information transfer calculations.

Also, the persistence parameters τ are only significant when the values of their corresponding S parameters are reasonably large. Taken together with the above comments on FDR, this means that the effective number of free parameters in the fit is somewhat unclear, but considerably less than 9.

### 2.4 *Simulations*

Under physiological conditions, cells are not exposed to voltage-clamp and respond stochastically rather than deterministically. For measurements of information transfer in spiking neurons, we therefore used postsynaptic conductance changes rather than EPSCs, and extended the model to provide a stochastic description of vesicle release. We first describe the use of conductances, and then stochastic release.

The release of a single vesicle produces a quantal postsynaptic conductance g. The maximum conductance induced in a postsynaptic cell in response to a single presynaptic AP is therefore $G_{SE} = n \cdot g$. The resulting peak EPSC amplitude is $A_{SE} = G_{SE}(V_m - E_{rev})$, where $V_m$ is the membrane potential in the postsynaptic cell and $E_{rev}$ is the reversal potential of the synaptic conductance. In a *deterministic* model, the resulting total synaptic conductance $G_{syn}$ would therefore be

$$G_{syn} = \sum_{\text{all connections}} G_{SE} U_{SE} P_V, \qquad (12)$$

in analogy with $I_{VClamp}$ (**Eq. 6**). When the postsynaptic cell is voltage clamped, $A_{SE}$ and $G_{SE}$ thus differ only by a multiplicative constant $(V_m - E_{rev})$, as do $I_{VClamp}$ and $G_{syn}$.[2]

In an unclamped cell, $V_m$ varies dynamically as the membrane conductance is altered, producing APs in the postsynaptic cell. To simulate this process, we fed $G_{syn}$ into a typical integrate-and-fire model cell:

$$\frac{dV_m}{dt} = \frac{1}{\tau_m} \left[ G_{syn}(t) \cdot R_N \cdot (E_{rev} - V_m) + (V_{rest} - V_m) \right]. \qquad (13)$$

$G_{syn}$ generates a current, which causes a change in membrane potential $V_m$ in the postsynaptic cell. Both the size of this current and the resulting change in $V_m$ depend upon $V_m$ itself. $V_{rest}$ here is the resting membrane potential, which we set to –65 mV, in line with that observed during the cortical 'up'-state [30-32] (as we assume that a single pair is most relevant to information transfer when the network is in an excited state). When $V_m$ passed threshold, we stepped it to 40 mV for 1 ms, after which we immediately hyperpolarised and allowed it to relax back to $V_{rest}$. We used a threshold for spike generation of –55 mV, a hyperpolarisation potential of –75 mV and a reversal potential of $G_{syn}$ of 0 mV, in line with the known properties of AMPA receptors [2]. Our choice of the integrate-and-fire model was based on the fact that whilst this model does not describe the voltage time courses in between action potentials accurately, it replicates the series of action potentials very well [33]. To mimic ligand unbinding from AMPA receptors, we also assumed conductances to decay exponentially over 2 ms. Changes in membrane potential occur exponentially with characteristic time $\tau_m$, which reflects the resistive and capacitive function of the cell's lipid membrane. We set $\tau_m$ to 50 ms, as we observed in neurons in layer IV (not shown). Similarly, we set the input resistance ($R_N$) of the model cell to 250 MΩ, following previous results from layer IV [9]. We integrated **Eq. 13** using a variable time step, typically far smaller than the resolution at which we calculated synaptic model evolution and input/output spike trains (1 ms).

---

2   The reader should note that **Eq. 12** is included only for the sake of explanation, and is superseded by **Eq. 15** in our analysis; we do not actually *employ* **Eq. 12**, nor the unclamped, deterministic model it implies, at all in this paper.

Next, we made the model fully stochastic. To do this we allowed vesicle release following comparison of $U_{SE} \cdot P_V$ with a random number between 0 and 1. Here, $n$ can only be 1. Following successful vesicle release, $P_V$ is set to zero and relaxes back to 1 with persistence $\tau_{VDD}$. **Eq. 1** therefore becomes

$$\frac{dP_V}{dt} = \frac{1 - P_V}{\tau_{VDD}} - P_V \delta(t - t_{success}), \tag{14}$$

where $t_{success}$ are the times at which release has been successful. **Eq. 12** then becomes

$$G_{syn}(t) = G_{SE} \delta(t - t_{success}). \tag{15}$$

**Fig. 4** illustrates the validity of our stochastic implementation. For this example, we have reverted to a voltage clamp configuration, in order to illustrate the correctness of our stochastic model by comparison with the output of the familiar deterministic, voltage-clamped model (Sec. 2.2). Response moments have been estimated in Tsodyks-Markram type models by the simulation of individual release events previously [15,34], although [15] contained some errors in its mathematical formulation.

## 2.5     *Information transfer by spike timings.*

At its simplest level, the information content of a dataset is determined by how many possible distinct values any dataset of its type can have, and how likely each of these is to occur. This is precisely the definition of the entropy of a particular type of dataset. To quantify the information contained in responses of our stochastic model, we used the direct method of measuring information transfer [18,19,35]. In this method, the mutual information [35]

$$I(R,S) = H(R) - H(R|S), \tag{16}$$

between stimuli *S* (presynaptic spike trains) and responses R (postsynaptic spike trains) is determined explicitly from the contents of the trains. H(R) here represents the total entropy (information) of the responses, whilst H(R|S) is the conditional entropy of the responses given a certain stimulus. These are given by

$$H(R) = -\sum_j p(r_j) \log_2 p(r_j), \tag{17}$$

$$H(R|S) = -\sum_i p(s_i) \sum_j p(r_j|s_i) \log_2 p(r_j|s_i). \tag{18}$$

The probability of response $r_j$ occurring is $p(r_j)$, and the probability of stimulus $s_i$ being presented is $p(s_i)$. The probability of response $r_j$ occurring *if stimulus $s_i$ is presented* is $p(r_j|s_i)$. *H(R)* is estimated by counting how often different response sequences occur over the entire range of stimuli, whereas *H(R|S)* is determined by considering the occurrences of different responses given only one specific stimulus. *H(R)* provides a measure of the possible information contained in a response. *H(R|S)* indicates how much of that information is simply random variation in *R* due to the inherent unreliability of the information channel, uncorrelated with *S* (i.e. noise).

Stimulus and response spike trains are digitised with a bin width Δt, where "1" denotes the presence of a spike in the bin, and "0" no spike. To limit response strings to a computationally manageable number of possible values, they are split into non-overlapping windows of length *T*. Each window contains a "word" consisting of *T*/Δt bits. Occurrences of different words are tallied over all windows in the set of responses, and counts are normalised to give $p(r_j)$, from which *H(R)* is determined via **Eq. 17**. We used Δt = 4 ms, giving a maximum measurable information transfer rate of $\Delta t^{-1}$ = 250 bits/s. We chose *T* = 16, 20, 24, 28, 32, 40, 48, 56, 68 and 80 ms, then used the variation of *H(R)* with window length to extrapolate to *T* = ∞ (see below).

The total information content of a stimulus ensemble $I(S)$ is determined in exactly the same way as $H(R)$, using **Eq. 17** with $s$ in place of $r$ and counting probabilities of the different stimuli $s_i$. We used randomly-interleaved 5 s Poisson stimuli with central frequencies of 10, 20, 30, 40 or 50 Hz, producing $I(S) = 121.1 \pm 3.1$ bits/s ($2\sigma$, 30 repetitions).

Conditional (noise) entropy $H(R|S)$ is calculated by determining the entropy of responses to a repeated stimulus sequence. We presented the same Poisson stimulus sequence multiple times, using the same counting method as for $H(R)$ to determine word probabilities. We did this with one Poisson realisation for each frequency (i.e. 10 Hz, 20 Hz, etc).

We used overlapping windows in determining $H(R|S)$, as noise in each bin is statistically independent (so noise is no more correlated in windows differing by a single position than in non-overlapping windows). This allows better sampling of information measures with a smaller number of responses. We measured probabilities $p(r_j|s_i)$ individually in each window position rather than the response train as a whole, using counts of words appearing in the same window across stimulus repeats. Because conditional entropy is a measure of variability in response to a given signal, different sections of the input stimulus constitute different stimuli $s_i$ to be summed over. As the probabilities $p(s_i)$ of different stimuli were all equal in our case, the outer sum in **Eq. 18** simply become an average over all stimulus frequencies and window positions.

To make any reasonable estimate of response probabilities in **Eqs. 17 and 18**, very many stimulus-response pairs must be analysed. We generated ~5.4 h of synthetic data for each combination of synaptic parameters, using ~2.7 h each to determine $H(R)$ and $H(R|S)$. We performed the extrapolations to infinite data size and window length considered a standard part of the direct method. We give an example of the latter in **Fig. 5A,** showing that the effect of using a finite window size is minimal. Very little downturn due to undersampling is seen with increasing window length, indicating that the extrapolation to $T = \infty$ is very reliable.

To determine the uncertainty of our information transfer measurements, we repeated measures 30 times each for four sets of synaptic parameters. Standard deviations of the resulting transfer rates are shown in **Fig. 5B**. The standard deviation is approximately proportional to the mean transfer rate (fitted line, $r = 0.987$, $p_{Pr} =$

0.013). We used the values of this fit as a proxy for the uncertainty in transfer rates in general: the estimated value is 8.0% (2σ). In the interests of computational expediency, we assumed that the variability of information measures does not depend explicitly upon the dynamic character of the synapse, only how well it can pass information. We do not expect this to necessarily be true in general, but the goodness of fit in Fig. 5B indicates that it is a reasonable approximation. The data points for spike timing presented in Figs. 6–9 are hence based on single measurements, with errors approximated as twice the benchmark standard deviation corresponding to that information value.

For connected pairs, we set $G_{SE}$ such that a postsynaptic spike would always be produced if vesicle release from the presynaptic cell was successful, and the membrane potential of the postsynaptic cell was at its resting value. This constitutes the requirement that every presynaptic spike has the *chance* of eliciting a postsynaptic spike; anything less would result in no postsynaptic spikes and no information transfer. We chose $G_{SE}$ = 30 nS to achieve this with our chosen integrate-and-fire parameters. Such a high quantal conductance was only required because of the very short time over which we applied the excitation (i.e. just one timestep at the simulation resolution of 1 ms) in comparison to the membrane time constant (50 ms). Alternative approaches would have been to apply a smaller conductance for longer, and/or to allow the input resistance and membrane time constant of the model postsynaptic cell to change with input activity. As we were only concerned with making every vesicle able to generate a postsynaptic AP, not the detailed influence of postsynaptic properties, these approaches are equivalent for our purposes.

Although this 'one vesicle, one spike' scenario is not physiologically realistic, reports exist of a single connection between neurons generating spikes at some synapses [36,37], or very small numbers of connections at others [38]. This may also be so during neuronal excited states such as the cortical 'up' state [30-32], or where dendritic amplification is strong [39]. With this value of $G_{SE}$, we can estimate the contribution of short-term plasticity to information transfer at single connections, and its influence upon spike-timing information, without needing (computationally horrendous) dynamic network simulations along with spike-timing measures. This provides a more realistic measure of information transfer than just considering EPSC timings, as the mechanism of spike generation and the interplay of $V_m$ with conductance changes introduces a further variability in resultant spike timings that is not present in EPSC timings.

### *2.6  Information transfer by spike rates.*

To analyse information carriage by spike rates alone, we considered two cases: the original configuration of a singly connected pair, and a simple toy model of a neural network, with 1500 presynaptic neurons connecting to one postsynaptic cell. This number of connections roughly corresponds to the ~$10^4$ synapses observed to impinge upon pyramidal cells [40], with about six synapses belonging to each individual connection [9,41]. We allowed all 1500 presynaptic cells to fire independently, and each connection to undergo dynamic changes in efficacy independent of the others. Our goal with this network is to investigate a very simple example of multiple connectivity, to see whether modulation of the information-passing characteristics of the network by short-term plasticity differ substantially from those of the single pair. Investigating the complex influences of short-term plasticity on larger and more realistic networks is beyond the scope of this paper, but when our results are combined with recent efforts in this direction [42], they should go some distance to making such work possible in the future.

In each case, an adapted version of the direct method was required to limit consideration to spike rates, and deal with the resulting computational and sampling constraints this imposed on information measurements. A summarised comparison of the different information measurement techniques we use in this paper is given in **Table 1**, contrasting the timing and rate measures in pairs and the rate measures in the network.

Our measures of rate information are of relatively low temporal and count resolution, and are probably undersampled. We made no effort to adjust for finite data size or word length, as data sizes and word lengths were not large enough to reasonably justify such extrapolation. Our measures are also prone to saturation and underspiking (described below). These systematic effects should not have influenced our qualitative conclusions, however. Our rate measures should thus be considered approximate estimations of information transfer by a rate code, more useful for their relative than absolute values.

For both rate measures, we considered the total number of postsynaptic spikes in every 500 ms of postsynaptic response. In order to deal with the added computational complexity of multiple spikes in each temporal bin (rather than just one or zero spikes per bin), we used stimuli of 2 s duration. This provided four non-overlapping bins per response, with a single window extending over the whole 2 s of the response. The combination of counts in these four bins constituted a word. We estimated response probabilities as normalised counts of the occurrence of each word, and used them to calculate $H(R)$ as per **Eq. 17**. Similarly, we calculated $H(R|S)$ with **Eq. 18,** using normalised counts of responses to repeated presentations of the same stimulus.

We used randomly interleaved 10, 20, 30, 40 or 50 Hz Poisson trains as stimuli, but this time with a step from one average frequency to another after 1 s, producing 25 possible average frequency combinations (or stimulus 'classes'; this included steps to the same frequency). We designed these stimuli specifically to test the abilities of the network to distinguish between different mean spike rates. As a rate measure considers only mean responses over large windows (500 ms bins in our case), it cannot measure the ability to distinguish different spike rates without an explicit rate change in the stimulus.[3] We did not repeat the same Poisson trains to presynaptic neurons for the purpose of determining $H(R|S)$, as the information measure is not sensitive to the resultant spike timing; we therefore simply drew repeats from the same stimulus class.

For rate measures on pairs, we generated ~5.4 h of synthetic data. We limited spike counts to 50 per bin (100 Hz), resulting in a maximum measurable information rate (**Eq. 17**) of $0.5 \cdot \log_2 51^4 \approx 11.3$ bits/s. The theoretical information content of the stimulus follows from the 25 possible stimulus classes presented over 2 s, giving $I(S) = 0.5 \cdot \log_2 25 \approx 2.3$ bits/s. We measured the stimulus information directly as $I(S) = 2.19 \pm 0.01$ bits/s ($2\sigma$, 30 repetitions). This is consistent with the theoretical prediction, considering that undersampling dictates that practical measures should slightly underestimate the true information content.

---

3 When spike timing is included in the information measure, the ability of synapses to distinguish between different spike rates is implicitly measured by virtue of using a stochastic stimulus sequence, even if the stimulus has a constant mean rate.

We estimated errors from 20 repetitions of a single information transfer simulation, giving $\overline{I(R,S)}$ = 0.512 bits/s and $\sigma_I$ = 0.018 bits/s ≈ 3.5% of $\overline{I(R,S)}$. We assumed that standard deviations of rate measures follow a similar relationship as seen in **Fig. 5B**, producing an estimated error of 6.9%. We chose less detailed error estimates with spike rates than with spike timings because of the low resolution of the rate measures.

Given the additional overhead in computing 1500 model synapses, we made further approximations to measure information carriage in the network. We presented 25 different instances of each of the 25 stimulus classes to presynaptic cells, so each calculation of $H(R)$ drew upon ~21 min of postsynaptic responses (instead of the ~2.7 h generated for pairs). We presented statistically independent trains to different presynaptic cells, meaning that for each ~21 min of postsynaptic response, we computed ~3 weeks of stimuli and actual synaptic behaviour. We generated as much data again to determine $H(R|S)$. We estimated errors from the standard deviation of 20 repetitions of a single measure. This gave $\overline{I(R,S)}$ = 1.258 bits/s and $\sigma_I$ = 0.042 bits/s ≈ 3.4% of $\overline{I(R,S)}$, leading to an estimated error of 6.8%. We sorted spike counts into 5 possible 'count bins', corresponding to counts of <15, 15–24, 25–34, 35–44 and >44. Using the combination of counts from the four 500 ms temporal bins, this allows $5^4$ = 625 possible responses. By **Eq. 17**, the largest measurable rate of information transfer in this coding scheme is 0.5·$\log_2$625 ≈ 4.6 bits/s. Because stimuli were generated by many presynaptic cells, measuring the total $I(S)$ was difficult. However, given that the stimulus is the same as in the measurement of rate information with the pair, we know its theoretical content to be ≈ 2.3 bits/s. This is consistent with our results, which approach but never exceed this value. Using this coding scheme, we measure the information content of stimuli given to individual presynaptic cells as $I(S)$ = 0.932 ± 0.084 bits/s (2σ, 30 repetitions). The coarseness of the rate binning clearly reduces the stimulus information content when only one presynaptic cell is considered, but the redundancy in the network compensates for this and recovers much of the true stimulus information.

In the network, the quantal input conductance of each synapse ($G_{SE}$) becomes a spike-rate normalisation factor, as output spike rates scale directly with $G_{SE}$ for any given number of presynaptic cells. Given the binning of rates in this measure, we need to carefully choose $G_{SE}$ to ensure that output spike rates were in the range of values that can be decoded by our chosen rate code (mostly between 15 and 44 counts per temporal bin). We scaled $G_{SE}$ to 700 pS to elicit comparable spike rates in the postsynaptic cell as in the stimuli. This value was the best compromise across the synaptic parameter ranges we considered. This constant scaling over different synaptic parameter values is our alternative to Zador's [21] input rescaling method (see *Discussion*), and what ultimately allows us to elucidate the effects of dynamic range upon rate information in a network with dynamic synapses (**Fig. 9A**). This is because changing the conductance scaling is equivalent to altering the dynamic range of the recipient network, as globally raising or lowering output spike rates is identical to increasing or decreasing the limits upon the bins into which spike counts are placed.

## 3    Results

### *3.1    Paired recordings and synaptic modelling*

Our sample of connections exhibited the full range of short-term dynamics previously seen in neocortex. We fitted experimental traces from our 11 pairs with the 9 parameters of the model synapse. We identified the connections as depressing ($N = 6$) or facilitating ($N = 5$) according to the optimised parameter values. Fits were very good, returning a mean reduced-$\chi^2$ value of 0.183, with a standard deviation of 0.174. We used these values (**Table 2**) to ensure that our subsequent estimates of information transfer were based on biologically realistic parameter values.

In earlier work [3,9], we used functional descriptors $R_{FDR}$ and $R_D$ to quantify the degree of FDR and the release-dependency of depression. The typical depressing connection in cortical layer IV/V documented in **Table 2** produces responses with $R_D = 0.45$, in close agreement with the $0.45 \pm 0.35$ found earlier in layer V, and $0.30 \pm 0.29$ found in layer IV. The typical depressing connection in **Table 2** has $R_{FDR} = 1.09$, compared with $1.37 \pm 0.25$ seen earlier in layers IV and V. The difference is roughly one standard deviation, and the typical value here is well within the range $0.9 < R_{FDR} < 1.9$ observed earlier. The range of values presented in **Table 2** for the depressing connection contains the previous range, permitting $R_{FDR}$ of between 1 and 3 ($R_{FDR}$

< 1 can only occur due to recording noise). The typical facilitating connection in **Table 2** produces $R_D = 0.84$ and $R_{FDR} = 1.07$.

### 3.2     *Information transfer by spike timings.*

We investigated how neuronal information transfer depends upon the strengths and persistences of VDD, RID, FAC and FDR, using our stochastic model synapse to produce a massive number of simulated postsynaptic spike trains, and the information theoretic 'direct method' [18,19,35] to analyse their information contents. The method explicitly permits information carriage by both individual spike timings and overall spike rates, but the former tends to dominate because it can carry more information.

Higher initial release probability ($U_0$; **Fig. 6A**) increases information transfer at both depressing and facilitating connections, but has a greater effect on depressing synapses. When RID and FDR are minimised at depressing connections to focus solely upon VDD, a slower recovery from VDD ($\tau_{VDD}$) has the opposite effect (**Fig. 6B**). Higher initial probabilities of release produce a greater reduction with increasing $\tau_{VDD}$, but transfer is higher overall in connections with higher initial release probabilities. Information transfer at facilitating connections depends upon the strength ($S_{FAC}$) and persistence ($\tau_{FAC}$) of facilitation, as shown in **Fig. 6C**, where RID, VDD and FDR were minimised. Stronger and longer facilitation increase information transfer. Stronger (larger $S_{RID}$) and longer RID (larger $\tau_0$) decreases information transfer at depressing connections (**Fig. 6D**; VDD and FDR minimised).

Information transfer is better in the presence of stronger and more persistent FDR (larger $S_{FDR}$ and $\tau_{FDR}$), both at facilitating (**Fig. 6E**) and depressing connections (**Fig. 6F**). A greater increase is evident between low and intermediate values of $\tau_{FDR}$ than between intermediate and high values.

Further interplay between VDD, RID, FAC and FDR is presented in **Fig. 7**. Consistent with the previous figure, stronger FAC increases information transfer at facilitating connections, whereas stronger RID decreases it (**Fig. 7A**). When FAC and RID are of similar strengths ($S_{FAC} = S_{RID}$), more complex effects occur. Synapses with $S_{FAC} \approx S_{RID}$ might be particularly useful in a computational sense, especially if combined with neuromodulation, as information transfer becomes highly sensitive to small variations in synaptic efficacy. More persistent FAC and FDR also both increase information transfer (**Fig. 7B, C**), whereas more persistent RID reduces it (**Fig. 7C**). Persistent VDD and RID decrease information transfer at depressing connections (**Fig. 7D**), but VDD has the greater effect.

Our results show a consistent influence of short-term plasticity upon of information transfer with spike timings: processes that increase the probability of neurotransmitter release increase the rate of information transfer, and processes that decrease release probability reduce information transfer. This agrees with results using static synapses [21].

### 3.3    *Information transfer by spike rates.*

To check whether this result holds in general, we investigated short-term plasticity in the context of information carriage by mean spike rates alone (**Fig. 8**). We used both a connected pair (dashed lines; filled symbols) and a network of 1500 presynaptic neurons contacting a single postsynaptic one (solid and dotted lines; open symbols). To consider just spike rates, and provide for the network configuration, we altered the standard direct method, resulting in three distinct neural information measures (**Table 1**). In this case, we specifically chose stimuli with no temporal information except for changes in the mean spike rate, producing stimuli with lower information contents than when spike-timing information was included. The network configuration was designed to operate under a pure rate-coding scheme, with entirely asynchronous inputs.

The relationships between synaptic plasticity and information transfer by a rate code in the pair (**Fig. 8**) are virtually identical to those seen with a spike-timing code (**Figs. 6 and 7**). Higher initial release probability enhances information transfer by both depressing and facilitating synapses, with greater effects seen at depressing connections (**Fig. 8A**). VDD (**Fig. 8B**) and RID (**Fig. 8C**) decrease information transfer, whilst FAC (**Fig. 8C**) and FDR (**Fig. 8E, F)** increase transfer rates.

In the network, increases in the probability of release mostly *decrease* information transfer with a rate code, with higher transfer rates achieved by the depressing than the facilitating synapse (**Fig. 8A**). VDD has little effect (**Fig. 8B**), but facilitation actually decreases information transmission (**Fig. 8C**). FDR can have positive or negative effects (**Fig. 8E, F)**, depending upon the synaptic parameters. The influence of plasticity upon rate information clearly differs between the pair and the network configuration; RID is the only process that has a similar effect upon both (**Fig. 8D**).

### *3.4    The influence of dynamic range.*

Why do these differences occur? The first hint is in the effect of facilitation upon rate information in the network (**Fig. 8C**): too much facilitation reduces information transfer to zero. In this case, strong facilitation causes all postsynaptic spike trains to contain more than 44 spikes per 500 ms counting bin, regardless of the frequency of presynaptic stimulation. By the definition of information [35], the postsynaptic spike rate then carries no information about the presynaptic rate, as all responses automatically fall into the same counting bin (refer to **Methods** for further information). This behaviour is a feature of a network where postsynaptic responses are not dynamically matched to the presynaptic effects of short-term plasticity; neither the decoding scheme nor the synaptic conductance of the postsynaptic cell are modulated to account for the level of presynaptic activity. This is just one way of imposing a finite dynamic range upon a system. If the decoding scheme were able to distinguish between many very high spike rates, or the spike rates of the postsynaptic cell were downscaled by e.g. modulation of intrinsic excitability, then the information in the presynaptic rate would be recoverable even in the presence of strong facilitation.

We went on to investigate how the expansion or contraction of postsynaptic dynamic range alters the influence of facilitation upon information transfer. We carried out tests for all three configurations already discussed: rate information in a network (**Fig. 9A**), rate information in a pair (**Fig. 9B**), and spike-timing information in a pair (**Fig. 9C**). We necessarily manipulated the dynamic range in different ways for each information measurement scheme. In the network simulation, we gradually decreased the postsynaptic quantal conductance ($G_{SE}$), reducing average postsynaptic spike rates. This expanded the effective dynamic range, as responses began to take on rates that the postsynaptic decoding scheme could distinguish. For mean rate measurements on the connected pair, we gradually lowered the frequency $f_{max}$ above which any two firing rates were considered identical, reducing the effective postsynaptic dynamic range. In the spike-timing arrangement, we gradually increased the bin width (Δt) used to digitise spike trains, reducing the maximum number of spikes that the postsynaptic cell could discern per second, and thus its dynamic range. In this case, reducing the dynamic range shows the transition from a spike-timing to a rate code; for $\Delta t = 500$ ms, the information measure is effectively the same as what we used for mean rates, where we counted spikes in 500 ms bins. (Stimuli and some technical aspects of the measurement procedure remained different, so information transfer rates are not identical; refer to **Methods** for more information.)

In the network (**Fig. 9A**), information transfer initially increases with decreasing $G_{SE}$, as responses are brought progressively back into the functional dynamic range of the postsynaptic neuron, until they reach the theoretical limit imposed by the stimulus information content (dashed line). As spike rates are reduced further (smaller $G_{SE}$), higher values of $S_{FAC}$ no longer result in all responses falling into the highest counting bin (>44 counts per 500 ms temporal bin), and the positive influence of facilitation upon information transfer is recovered. As $G_{SE}$ and output rates are reduced even further, counts begin to instead build up in the *lowest* counting bin, giving the response a low information content.

Conversely, the positive gradient of information transfer with facilitation strength in the pair slowly reduces to zero, and then becomes negative with decreasing dynamic range (lower traces in **Figs. 9B** and **C**). The negative gradient at low dynamic range is due to precisely the same effect seen in the network: stronger facilitation causes more responses to exceed the maximum count rate discernible by the decoding scheme, making them indistinguishable from each other. The negative gradients of these lower traces are even more striking if viewed on a more appropriate scale, and with error bars (e.g. **Fig. 9D**). Thus, with pairs and a reduction in dynamic range, we recover the effects seen in a network.

Information carried by spike timing is not affected by dynamic range, as the crossover into the negative gradient regime in **Fig. 9C** only occurs when bin widths become sufficiently large that the code has effectively become a mean rate code. We have shown this unequivocally for a pair of neurons, but argue in the Discussion and Conclusions that it might hold even for a network.

The effects of dynamic range can also be used to explain the differing impacts of VDD and FDR on rate-coded information in pairs and in the network. When FDR is very strong (**Figs. 8E** and **F**), postsynaptic spike rates are quite high, and exactly the same negative effect upon transmission of rate-coded information in the network is seen as with strong facilitation.

The network's information-passing ability is unaffected by the timescale of recovery from VDD (**Fig. 8B**) because of its redundancy: the presence of 1500 presynaptic cells means that even when VDD is very long-lived, postsynaptic spike rates are still high enough to fall within the dynamic range of the target network. If spike rates were reduced by choosing a sufficiently small $G_{SE}$, more persistent VDD would decrease information transfer; likewise, a sufficiently large $G_{SE}$ would cause more persistent VDD to actually *increase* information transfer.

The redundancy of the network means that the size of the time windows over which rates are measured *in the network* should have relatively little qualitative impact on the observed influences of short-term plasticity, as long as output rates remain within the dynamic range of the target network, and the windows are still large enough for the neural code to be considered a rate code.

## 4      Discussion and Conclusions

We have presented a new model of presynaptic dynamics and obtained fitted values of its parameters from paired-cell recordings in neocortex. We used the range of fitted parameters from 11 connections to show how the information transferred between cells depends upon vesicle-depletion depression (VDD), release-independent depression (RID), frequency-dependent recovery (FDR) and facilitation (FAC). We showed that the influences of these four processes differ depending upon the nature of the neural code (either a rate or a spike-timing code) and the level of network connectivity (information transmission to a single postsynaptic cell or to an entire network). We went on to show that surprisingly, the influences of the choice of neural code and the degree of synaptic connectivity are in fact equivalent. This is because both influences arise purely from the effect of a finite postsynaptic dynamic range upon a rate code.

Our results indicate that information transfer is qualitatively affected by short-term plasticity in line with release probability, but exceptions occur when target cells possess a limited dynamic range and the neural code is based only on spike rates. This result is intuitively appealing, yet there has so far been no rigorous *a*

*priori* argument from our current understanding of synaptic plasticity and the mathematical definition of information [35] for it to necessarily be so. This insight could only be concretely gained through detailed simulations and information-theoretic measurements.

We generally observed a greater transmission of rate information by the network than the pair. This might be expected from the greater connection redundancy in a network: much of the information loss due to synaptic unreliability is compensated for by duplication of the original signal. Due to synaptic unreliability, rate information in a paired connection is compressed into differences between low rates of spiking, giving a lower spike count resolution and hence overall information rate. We see in **Fig. 9** that the effects of reduced dynamic range on a rate code become more pronounced in a network. This can also be understood in terms of the redundancy in the network: by compressing rates into a smaller dynamic range than is actually available, the pair is far better insured against contractions of the available range than the network, which utilises the entire range.

Our observation that a network is far more capable of conveying rate-coded information than a pair contradicts a finding with static synapses by Zador [21]. He found that increased connectivity results in poorer transmission of rate information. We believe this is primarily because he did not use an information-theoretic approach to quantify rate information, and performed a rescaling of input spike rates. On the other hand, we found that the effects of short-term synaptic plasticity upon spike-timing information typically tracks the effects upon release probability, which is highly consistent with Zador's spike-timing results (which *were* based on an information-theoretic analysis). He showed that information transfer with static synapses is a monotonically increasing function of release probability. This was despite his input rescaling method, which not only produced stimuli of varying information contents and effective degrees of synchrony, but demanded that the postsynaptic cell fired at a constant mean rate.

In light of our own results, the spike-timing findings of Zador may be robust to absence of an information-theoretic measure, and his use of input rescaling. If this it true, it suggests that our spike-timing findings at single pairs would hold also in networks, as working with static synapses gave Zador the computational resources to carry out his spike-timing measures in a partially synchronised network. Consider also our result that (in pairs at least), the influence of dynamic range only really seems to become relevant when rates dominate the neural code over spike timings (**Fig. 9C**). Together, these hint that the impact of short term plasticity upon spike-timing codes could be insensitive to the dynamic range of the recipient network in general, regardless of whether that is a single cell or many.

In considering the dynamic range over which a postsynaptic cell is able to 'decode' its spiking rate, we have deliberately ignored exactly what gives rise to this ability. It could be that the postsynaptic cell is tuned for spiking at certain frequencies by its membrane properties, or itself possesses synapses tuned for maximal response to certain frequencies [8,13,16]. As such, what is truly important is the overall dynamic range of the whole network a synapse feeds into, as this dictates the effective dynamic range of any particular component within it.

We observed consistent effects from VDD, RID, FAC and FDR upon neuronal information transfer throughout the entire set of parameter values obtained from our sample of neocortical synapses, including how these effects change under different coding regimes and decoding abilities of the recipient networks. Because the effects exist robustly across our broad parameter set, we predict that they will also hold in other brain regions.

Admittedly, finite computational resources meant that some aspects of our simulations were not optimal. Our information measures were based on either a large network of 1500 presynaptic cells contacting a single postsynaptic cell, or a single presynaptic cell contacting a postsynaptic cell via a single release site. Realistic connections typically contain more than one (but often far less than 1500) presynaptic cells, each with multiple release sites. Our rate-coding information measures were based on relatively large temporal bins (500 ms), producing rather coarse and probably undersampled information measures. These aspects should be improved upon in later works, and their impact on results carefully examined. We are nonetheless confident that our results capture the most important aspects of the influence of short-term plasticity upon neuronal information transfer.

The differential effects of dynamic range upon the four forms of short-term plasticity, and their dependence upon the size of the network and its coding strategy, provide neural networks with a series of orthogonal strategies for passing and processing information. This gives the brain the potential to perform tasks simultaneously over different coding regimes and network levels. To maximally realise this potential, efficient brains probably match the dynamic character of their synapses to processes known to influence dynamic range, such as neuromodulation and homeostatic plasticity [43], during development and behaviour.

**Acknowledgments:** We thank Bruce Graham, Greg Stuart and Stephen Williams for helpful discussions and comments on early versions of the manuscript.

# Figures

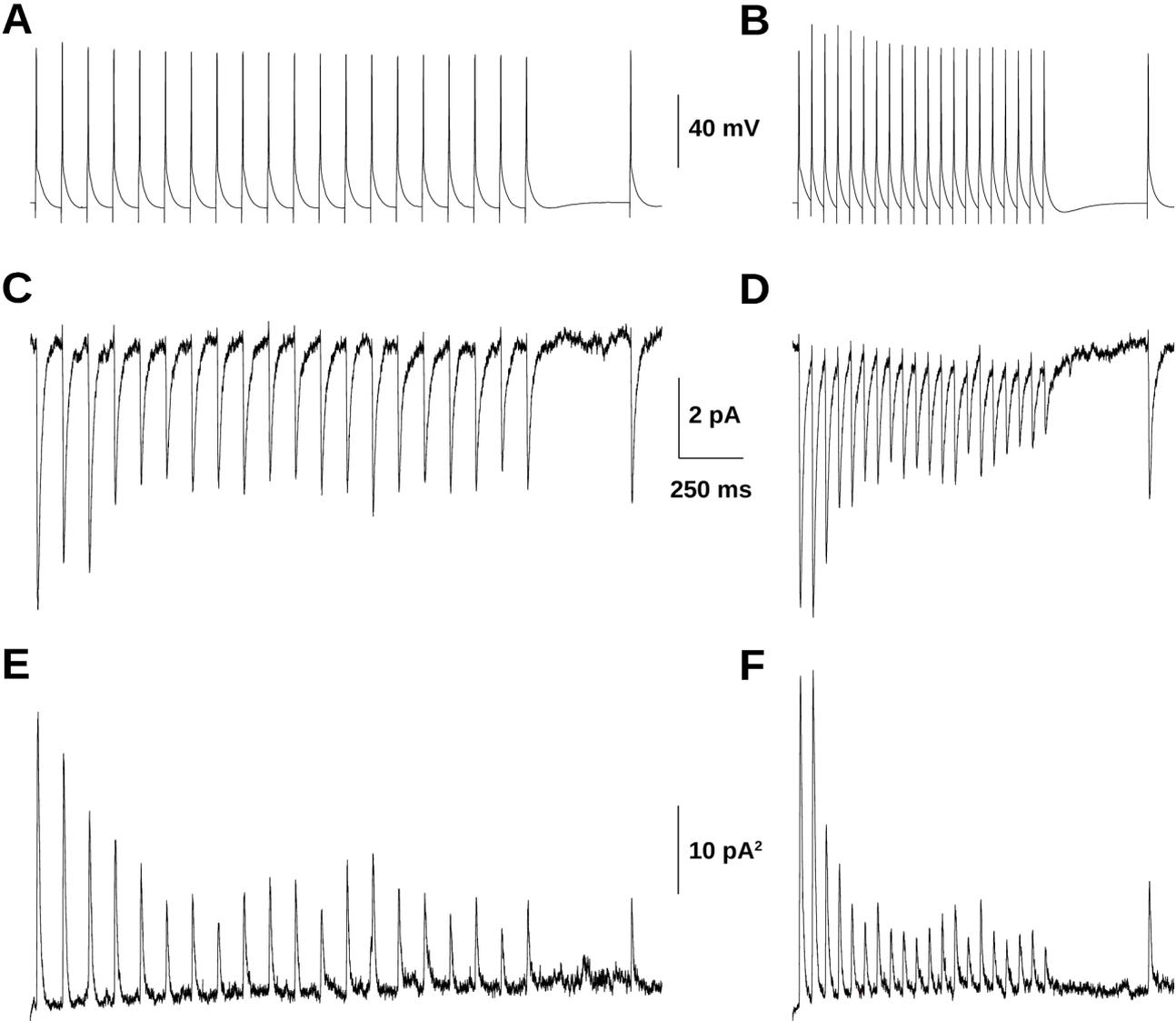

**Figure 1** Example recordings: presynaptic AP sequences at 10 **(A)** and 20 Hz **(B)**, with average EPSC **(C, D)** and variance time courses **(E, F)**.

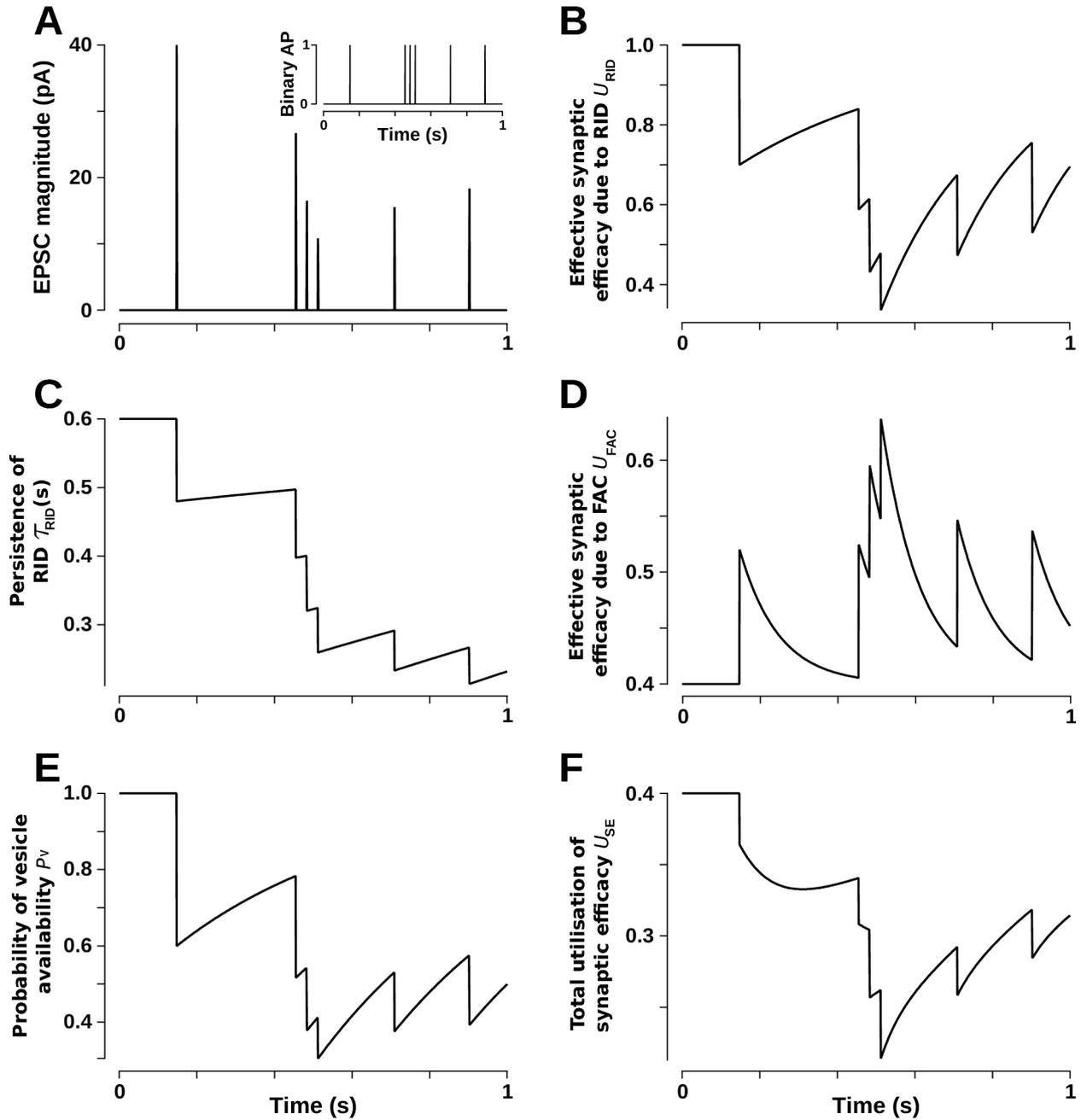

**Figure 2** An example of deterministic model evolution in response to a presynaptic train of APs. The stimulus is a randomly generated Poisson train centred at 10 Hz (**A**, **inset**), and final model output (**A**) is in the form of EPSC magnitudes. Plots show the time evolution of (**B**) the amount of release machinery free from inactivation by RID ($U_{RID}$), (**C**) the recovery time scale from RID ($\tau_{RID}$), (**D**) the degree of facilitation ($U_{FAC}$), (**E**) the probability of vesicle availability ($P_V$), and (**F**) the probability of vesicle release given availability ($U_{SE}$). Note the typical instantaneous drops or rises followed by exponential recoveries or decays in $\tau_{RID}$, $P_V$, $U_{RID}$ and $U_{FAC}$. More complex behaviour can be seen in the evolution of $U_{SE}$ (e.g. a bowl-shaped

dip following the first AP), reflecting the multiplicative relationship between $U_{RID}$ and $U_{FAC}$ that gives rise to $U_{SE}$ (**Eq. 5**). It is this feature of the new formalism that allows a model synapse to exhibit both facilitation and RID simultaneously, without interference. Parameters used: $A_{SE} = 100$ pA, $U_0 = 0.4$, $S_{RID} = 0.3$, $S_{FAC} = 0.2$, $S_{FDR} = 0.2$, $\tau_0 = 0.6$ s, $\tau_{VDD} = 0.5$ s, $\tau_{FAC} = 0.1$ s and $\tau_{FDR} = 2$ s.

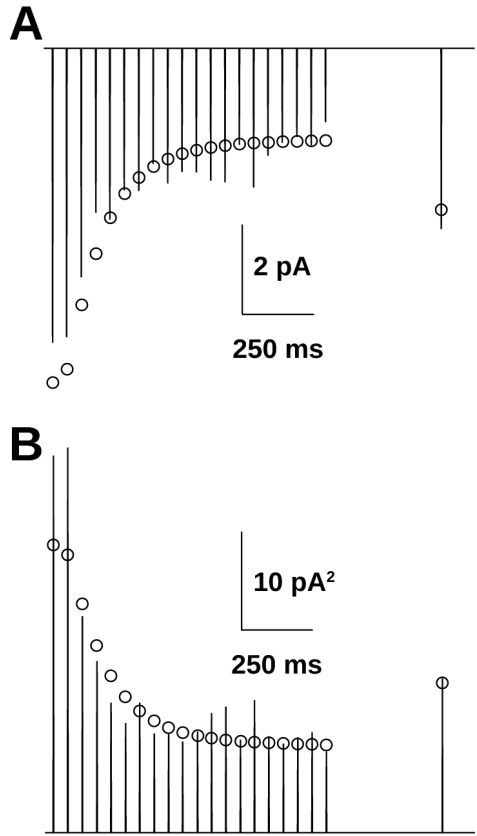

**Figure 3** Example fits: fitted model output (○) together with peak EPSCs **(A)** and variances **(B)**, shown for the 20 Hz stimulus. Peak EPSC and variance amplitudes were extracted from **Figs. 1D and 1F**, respectively. The fit was constrained by co-fitting responses to the 10 Hz stimulus. The model replicates experimental data very well; standard deviations are considerably greater than the discrepancy between theory and experiment (reduced $\chi^2 = 0.080$). Fitted parameters: $A_{SE} = 56.6$ pA, $U_0 = 0.13$, $S_{RID} = 0.22$, $S_{FAC} = 0.12$, $S_{FDR} = 0$, $\tau_0 = 0.45$ s, $\tau_{VDD} = 0.75$ s, $\tau_{FAC} = 0.06$ s and $\tau_{FDR} = \emptyset$. ("$\emptyset$" means that a persistence is irrelevant because its associated $S_{FAC}$ or $S_{FDR}$ value is zero.)

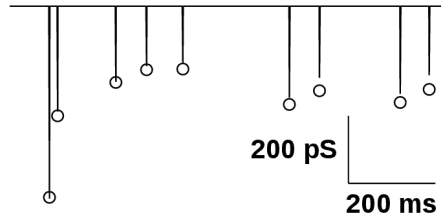

**Figure 4** Deterministic (trace) and averaged stochastic (○) model outputs. Stochastic averages were taken over 10 000 repeats. Our stochastic model clearly converges to the deterministic output. Parameters used are as per **Fig. 2**.

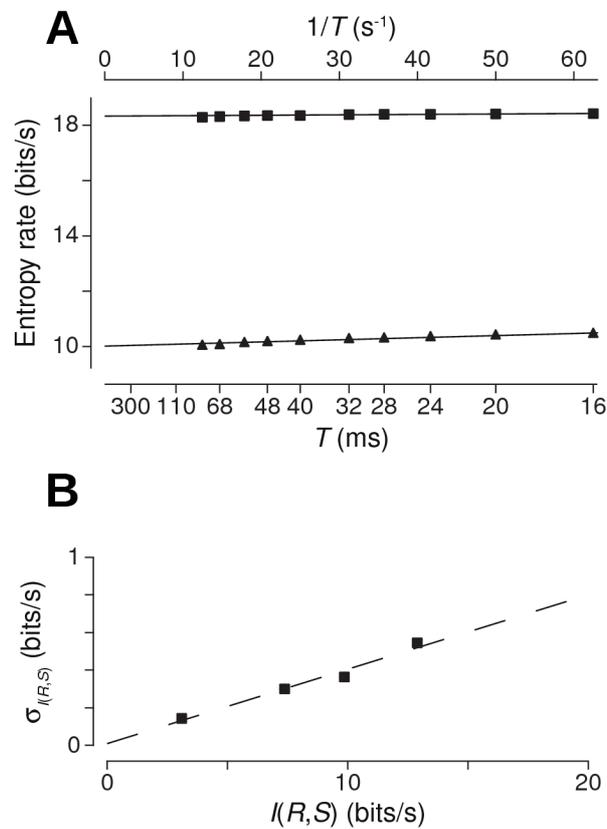

**Figure 5 (A)** An example extrapolation to infinite window length (*T*) as per the 'direct method'. The final transfer rate is the difference between the *y*-intercepts of the two extrapolations (■ total entropy, ▲ noise entropy), in this case 8.31 bits/s. Little falloff with window length is evident, suggesting that undersampling has a minimal effect upon our measures. **(B)** Estimates of errors in spike-timing information measures *I*(*R*,*S*). Estimates were derived from 30 repeated measures. The corresponding linear fit was 0.040·*I*(*R*,*S*) +

0.008 ($r = 0.987$, $p_{Pr} = 0.013$). Given the very small vertical offset, 8.0% ($2\sigma$) errors were assigned to measures in **Figs. 6 and 7**.

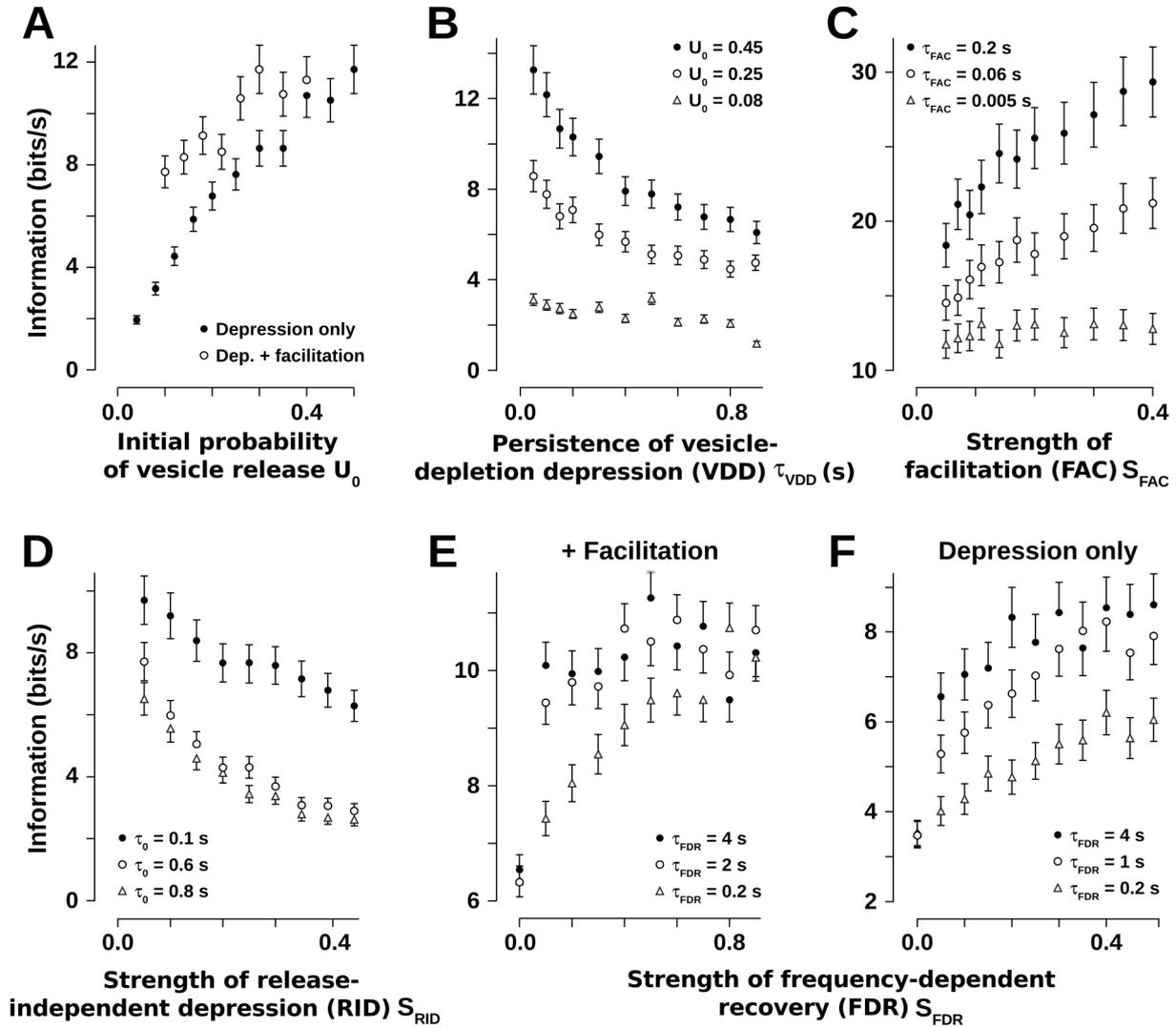

**Figure 6** Short-term plasticity affects information transfer with spike timings as it affects release probability. Error bars are $2\sigma$ in all plots except E, where $1\sigma$ is given for clarity. (**A**) Greater initial probabilities of release ($U_0$) increase information transfer. Other parameters are as per **Table 2**. Effects are more pronounced at depressing than facilitating connections. (**B**) More persistent VDD decreases transfer at depressing connections for different values of $U_0$, with greater initial release probabilities allowing a greater decrease. Parameters are modified to focus exclusively upon effects of RDD: $A_{SE} = 100$ pA, $S_{RID} = 0.1$, $S_{FAC} = 0$, $S_{FDR} = 0$, $\tau_0 = 0.6$ s, $\tau_{FAC} = \varnothing$ and $\tau_{FDR} = \varnothing$. (**C**) Stronger ($S_{FAC}$) and more persistent facilitation ($\tau_{FAC}$) increases

information transfer. Parameters are modified to focus exclusively upon effects of FAC: $A_{SE}$ = 100 pA, $U_0$ = 0.25, $S_{RID}$ = 0.15, $S_{FDR}$ = 0, $\tau_0$ = 0.15, $\tau_{VDD}$ = 0.05 s and $\tau_{FDR}$ = ∅. (**D**) Stronger ($S_{RID}$) and more persistent RID ($\tau_0$) decreases transfer at depressing connections. Parameters are modified to focus exclusively upon effects of RID: $A_{SE}$ = 100 pA, $U_0$ = 0.25, $S_{FAC}$ = 0, $S_{FDR}$ = 0, $\tau_{VDD}$ = 0.3 s, $\tau_{FAC}$ = ∅ and $\tau_{FDR}$ = ∅. (**E, F**) Stronger ($S_{FDR}$) and more persistent FDR ($\tau_{FDR}$) increase transfer at facilitating (E) and depressing (F) connections. Other parameters are as per **Table 2**.

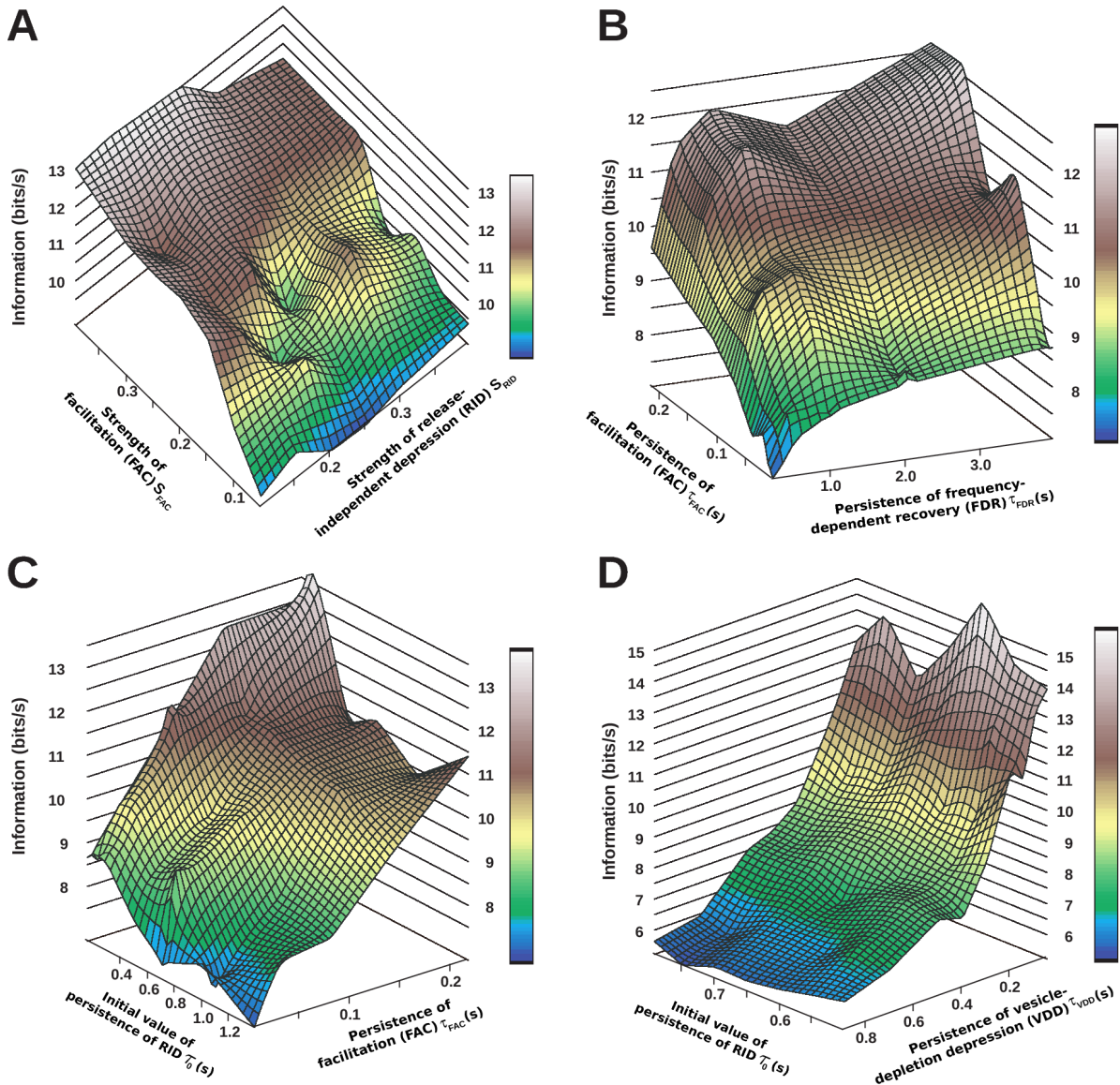

**Figure 7** (Colour Online) Extended investigations into the impact of short-term plasticity upon information transfer with spike timings. Error bars are omitted for clarity, but can be inferred as 8.0% of information values. All local variation in surfaces is below this level, except that seen in (**A**): the impact of the strength of RID ($S_{RID}$) and FAC ($S_{FAC}$) in a typical facilitating connection. RID clearly reduces information transfer and facilitation increases it, although a more complex interaction can be seen in the undulating region of the surface around $S_{FAC} = S_{RID}$ (see text). (**B**) Time courses of FAC ($\tau_{FAC}$) and FDR ($\tau_{FDR}$) in a facilitating connection. Both processes enhance information transfer if slower decays are allowed, but a greater difference is seen at lower values of $\tau_{FDR}$ than at higher values. (**C**) Time courses of FAC ($\tau_{FAC}$) and RID ($\tau_0$, the initial value of $\tau_{RID}$) at a facilitating connection, confirming the generally positive influence of facilitation upon transfer rates, and the negative influence of RID. (**D**) Time courses of RID ($\tau_0$) and VDD ($\tau_{VDD}$) in a typical depressing connection. Whilst greater persistence of each decreases information transfer, VDD can be seen to have the larger influence in this case. Except for those shown on the axes, all parameter values were held constant at the values defined in **Table 2**.

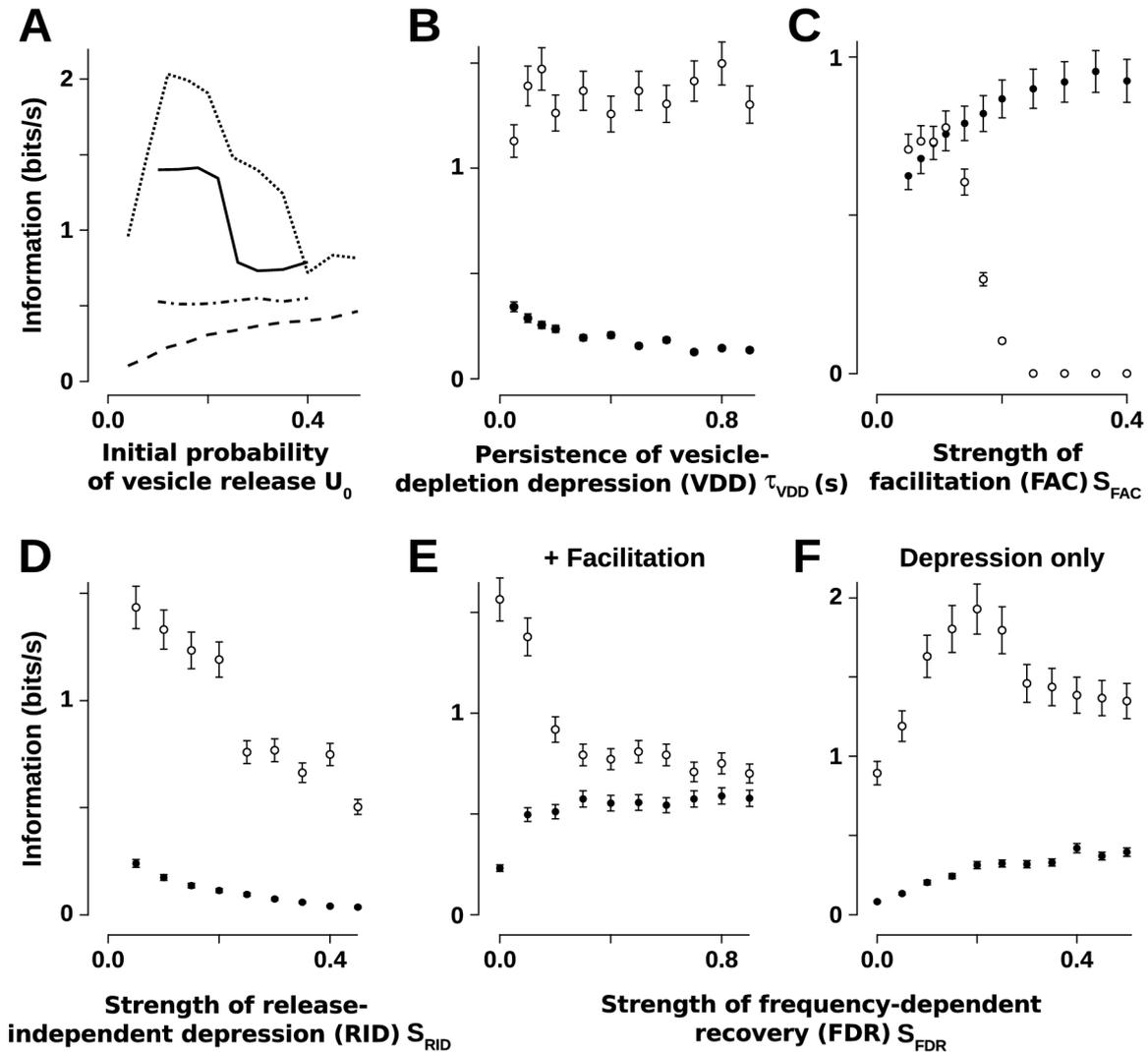

**Figure 8** Effects of short-term plasticity upon information transfer with spike rates in pairs (●) and networks (○). Error bars are 2σ. Parameter values in subfigures are as given for the central traces in equivalent parts of **Fig. 6**. (**A**) Information transfer is enhanced by higher initial release probabilities ($U_0$) in pairs (dashed and dot-dashed lines, depressing and facilitating connections, respectively), but not in a network (dotted and solid lines, depressing and facilitating connections respectively). (**B**) VDD decreases transfer between depressing pairs, but has no significant effect upon information transfer by the network. (**C**) FAC enhances transfer between facilitating pairs, but mostly hampers it in the network. (**D**) RID decreases information transfer by depressing pairs and the network. (**E, F**) Stronger FDR (greater $S_{FDR}$) increases transfer by both facilitating (E) and depressing (F) pairs, but has inconsistent effects in the network.

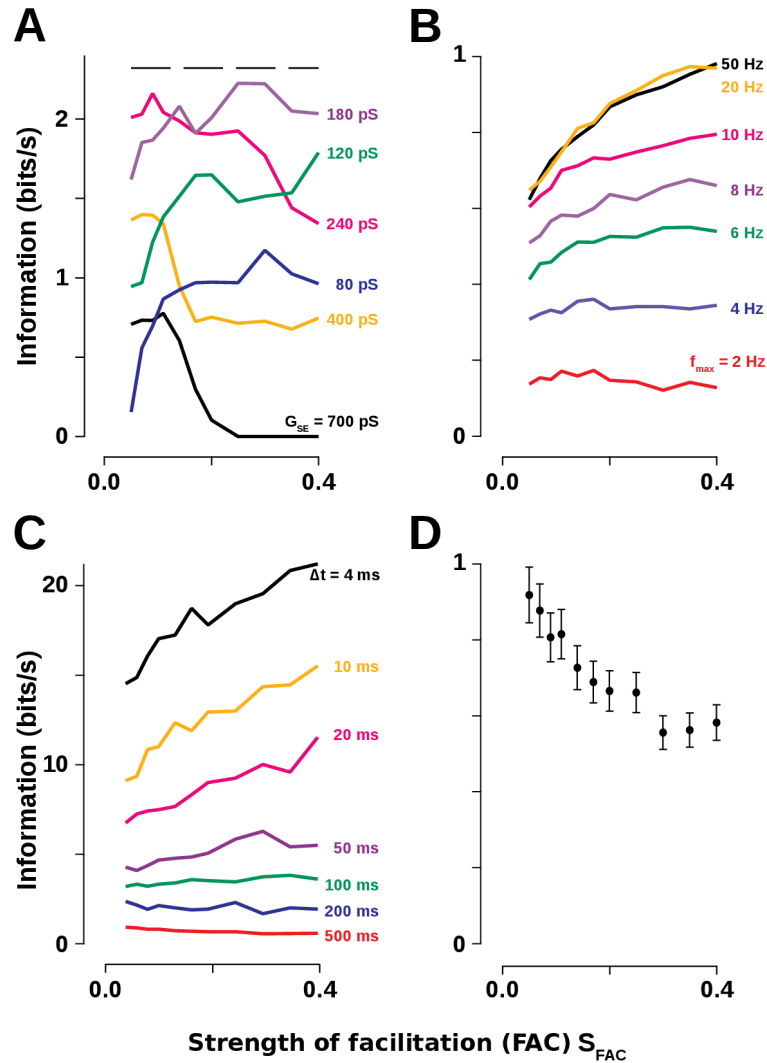

**Figure 9** (Colour Online) Effects of dynamic range upon information transfer with facilitating synapses. Error bars are omitted for clarity in A, B and C, but can be inferred as 6.8, 6.9 and 8.0% respectively. Parameter values are as for **Fig. 5C**. (**A**) Rate information in a network. The dynamic range of the postsynaptic cell was increased by reducing $G_{SE}$, recovering a positive gradient. (**B**) Rate information between a single pair. By decreasing the maximum frequency $f_{max}$ above which the postsynaptic cell can tell two frequencies apart, we reduced the dynamic range to the point where the gradient shows signs of becoming negative. (**C**) Spike timing information in a single pair. We steadily reduced the dynamic range of the postsynaptic cell by increasing the bin width $\Delta t$ in which spikes were detected, eventually recovering the negative gradient seen in subfigure A at low dynamic ranges in the network. (**D**) A close-up of the lowermost trace of subfigure C with error bars, clearly indicating that the negative gradient is significant. Despite using

the direct method with a single pair of neurons (as per **Figs. 3 and 4**), facilitation clearly reduces information transfer in this case, due to the reduced dynamic range induced by choosing Δt = 500 ms.

# Tables

**Table 1:**

|  | Pair, spike timing (direct method) | Pair, mean rate | Network, mean rate |
|---|---|---|---|
| $\Delta t$ | 4 ms | 500 ms | 500 ms |
| $T$ | 16 – 80 ms → ∞ | 2000 ms | 2000 ms |
| Bins / word | 4 – 20 → ∞ | 4 | 4 |
| Stimuli | 5 s, single freq. | 2 s, two freq. | 2 s, two freq. |
| Number of stimuli | ~2000 | ~5000 | 615 (× 1500) |
| $t_{total,stimulus}$ | ~5.4 h → ∞ | ~5.4 h | ~6 weeks |
| $t_{total,response}$ | ~5.4 h → ∞ | ~5.4 h | ~42 min |
| Spike counting | 0 / 1 | 0 – 50 | 0 – 50 in 5 bins |
| $I(R,S)$ | 250 bits/s | 11.3 bits/s | 4.6 bits/s |
| $I(S)$ | 121.1 ± 3.1 bits/s (2σ) | 2.19 ± 0.01 bits/s (2σ) | ~2.3 bits/s |

**Table 1** Summary of the 3 different neural information measures used. Measures corresponding to the 1st column are presented in **Figs. 6 and 7**, and to the 2nd and 3rd in **Fig. 8** (● and ○, respectively).

**Table 2:**

| Parameter | Depression only (N = 6) | | Depression + facilitation ($N$ = 5) | |
|---|---|---|---|---|
|  | Typical | Range | Typical | Range |
| $U_0$ | 0.25 | 0.04 – 0.50 | 0.25 | 0.11 – 0.36 |
| $\tau_{VDD}$ (s) | 0.50 | 0.30 - 0.90 | 0.50 | 0.05 – 0.75 |
| $S_{RID}$ | 0.25 | 0.10 – 0.40 | 0.18 | 0.15 – 0.22 |
| $\tau_0$ (s) | 0.60 | 0.57 - 0.77 | 0.30 | 0.15 – 1.40 |
| $S_{FDR}$ | 0.30 | 0 – 0.40 | ~0.20 | 0 – 0.90 |
| $\tau_{FDR}$ (s) | 1 / 2 | 0.20 – 2.00 | 2 | 0.20 – 4 |
| $S_{FAC}$ | 0 | 0 | 0.10 | 0.07 – 0.39 |
| $\tau_{FAC}$ (s) | ∅ | ∅ | 0.06 | 0.003 – 0.22 |

**Table 2** Typical fitted parameter values and ranges extracted from experimental recordings. (Typical in this case means modal when values are binned with a resolution smaller than about a fifth of the observed range). Connections are classified as depressing if $S_{FAC} = 0$, facilitating otherwise. The $\varnothing$ symbol is used to denote a parameter whose value is immaterial (or 'free'), if its value makes no difference to the synaptic model output. For example, the time scale of recovery from facilitation is immaterial if $S_{FAC} = 0$, since this means facilitation does not occur. The typical depressing value given for $\tau_{FDR}$ ("1 / 2") reflects the fact that data were inconclusive as to whether $\tau_{FDR} = 1$ or 2 s was more typical; $\tau_{FDR} = 1$ s was arbitrarily chosen for later simulations requiring typical depressing parameters.